\documentclass{PoS}
\usepackage{epsfig}
\usepackage{graphicx}

\title{Subtractive renormalization of chiral effective theory NN potentials
up to next-to-next-to-leading order}

\ShortTitle{Subtractive renormalization of the NN system}

\author{\speaker{C.-J.~Yang}, Ch.~Elster, and D.~R.~Phillips\\
        Institute of Nuclear and Particle Physics, and Department of Physics and
Astronomy, Ohio University,\\
Athens, OH 45701, USA\\
        E-mail: \email{cjyang, elster, phillips@phy.ohiou.edu}
        }

\abstract{
We have developed a subtractive renormalization method with which we can
evaluate nucleon-nucleon (NN) scattering phase shifts produced by the NN potential obtained at
leading, next-to-leading, and next-to-next-to-leading order (NNLO) in
chiral effective theory ($\chi$ET). In this method the low-energy constants
associated with short-distance NN physics are eliminated from the
Lippmann-Schwinger equation (LSE) for the NN t-matrix, in favor of physical
observables. This allows us to straightforwardly compute scattering phase
shifts for ultra-violet cutoffs of at least 10 GeV. We then perform detailed
analyses of the maximum cutoff at which the use of a $\chi$ET NN potential
in the LSE makes sense.

Specifically, we show that:

(a) our subtractive renormalization technique reproduces known results for
the LO potential, in both S- and P-waves;

(b) a parameterization of short-distance physics in the NNLO potential in
terms of an energy-dependent contact term creates scattering resonances and
shallow bound states in S-wave channels once cutoffs larger than 1 GeV are
considered;

(c) the more conventional momentum-dependent contact term in the NNLO
potential has problems of its own at cutoffs larger than 1 GeV;

(d) the NNLO potential yields P-wave phase shifts that have significant
dependence on renormalization point.

(e) for cutoffs smaller than 1 GeV, using spectral-function regularization
for the long-distance part of the potential produces results that vary with
the cutoff and depend on the renormalization point less than if dimensional
regularization is employed to compute the two-pion-exchange graphs.

Based on all these results we conclude that, once cutoffs larger than the
chiral-symmetry breaking scale are employed, iteration of the
two-pion-exchange piece of the $\chi$ET NN potential in the LSE does not
satisfy all of the criteria required for successful renormalization of the
problem.
}

\FullConference{6th International Workshop on Chiral Dynamics, CD09\\
		July 6-10, 2009\\
		Bern, Switzerland}
		
\begin{document}
\section{Introduction}
Chiral perturbation theory ($\chi $PT) is an effective field theory (EFT) that enables calculation in
the non-perturbative region of QCD. The use of EFT in nuclear\ systems
conveys two major advantages. First, we gain physical insight into the behavior of the strong interaction at large
distances, which is difficult to calculate {\it ab initio} from QCD. Second, our calculations become more accurate order by order, thus allowing systematic error
control. An EFT becomes most powerful when a clear and large seperation
between the low- and high-energy scale in the problem is possible. It has
been shown that $\chi $PT is quite successful in the low-energy ($<1$ GeV)
mesonic sector. In principle, the same theory should work in the low-energy
nucleon-nucleon (NN) sector as well.

However, it has been almost two decades since $\chi $PT was first applied to the
problem of NN system\cite{We90}, and difficulties still
remain. Standard $\chi$PT power counting, which would predict no bound state for the deuteron, does not apply to the NN problem,
because of infrared enhancements of the NN
interaction. A nonperturbative treatment of at least part of the NN interaction is thus a necessary ingredient. 
One needs to either iterate
the NN potential computed from $\chi$PT using the Lippmann-Schwinger equation (LSE)~\cite{Or96, Ka97, Le97,
Ep99,Ge99, ES01,EM02,Ol03,Ep05,Dj07,Be02,PVRA04A,PVRA04B, NTvK05,EM06,
PVRA06A,PVRA06B, En07}, or determine which part of the potential can be treated in perturbation theory~\cite{KSW96,
Fleming:1999bs,Fleming:1999bs, Be08, BMcG04,BB03,Bi06, LvK08}%
. So far there is no consensus as to which of these two alternatives is superior. 

Here, we adopt the former approach. In this ``chiral effective theory" ($\chi$ET) the
behavior of the $\chi$PT potential at high momentum necessitates that a cutoff $\Lambda $ be placed on the momenta in the LSE. It is
then natural to ask what values of $\Lambda $ can be used, if renormalization is to be successfully carried
out. Difficulties in answering this question have both a technical
part, i.e. it is hard to perform a fit for all unknown low energy
constants at high cutoffs because of \textquotedblleft
fine-tuning\textquotedblright; and conceptual problems regarding
what a successful renormalization is, e.g. is it sufficient that observables be (approximately) cutoff-independent? (See, e.g., Ref.\cite{EG09}, for a recent discussion.) 

In Sec.~\ref{sec-subtract} we outline a subtractive renormalization technique that solves the \textquotedblleft
fine-tuning\textquotedblright\ problem. This technique thus allows us to assess how well $\chi$ET at large cutoffs satisfies criteria for successful renormalization. 
We have used subtractive renormalization to calculate the NN scattering amplitude obtained by using leading-order (LO), next-to-leading-order (NLO), and NNLO $\chi$PT NN potentials \ in the LSE, for cutoffs up to $\Lambda=19$ GeV. We show some results of these calculations in Sec.~\ref{sec-result}, and examine the conditions under which $\chi$ET is really improved, order by order, after renormalization. We do this for both the dimensionally-regularized (DR) and spectral-function-regularized (SFR)~\cite%
{epsfr} $\chi$PT potentials, and consider both energy and momentum-dependent contact terms. More details regarding all these methods and results can be found in Refs.~\cite{Ya08,Ya09p,Ya09s}.

\section{Main ideas of subtractive renormalization}
\label{sec-subtract}
The main idea of our subtraction method is to construct the fully off-shell
partial-wave $t$-matrix from the knowledge of the long-range part of the
potential and the on-shell value of the $t$-matrix for zero energy~\cite{Ya08,Ya09p,Ya09s,HM99,AP04}.
The partial-wave LSE is given by
\begin{equation}
t_{l^{\prime }l}(p^{\prime },p;E)=v_{l^{\prime }l}(p^{\prime
},p)+\sum_{l^{\prime \prime }}\frac{2}{\pi }M\int_{0}^{\Lambda }\frac{%
dp^{\prime \prime }\;p^{\prime \prime }{}^{2}\;v_{l^{\prime }l^{\prime
\prime }}(p^{\prime },p^{\prime \prime })\;t_{l^{\prime \prime }l}(p^{\prime
\prime },p;E)}{p_{0}^{2}+i\varepsilon -p^{\prime \prime }{}^{2}}.
\label{eq:7}
\end{equation}%
Where $p_{0}^{2}/M=E$ is the center-of-momentum (c.m.) energy and $\Lambda $
the cutoff parameter. The incoming (outgoing) angular momenta
are indicated by $l$ ($l^{ \prime}$). The potentials are defined
as: 
\begin{equation}
v_{l^{\prime }l}(p^{\prime },p)=v_{l^{\prime }l}^{LR}(p^{\prime
},p)+C_{l^{\prime }l}p^{\prime }{}^{l^{\prime }}p^{l}f(p,p^{\prime};E),  \label{eq:vlprimel}
\end{equation}%
where $p(p^{\prime })$ indicates the incoming (outgoing) momentum in the
c.m. frame, $v_{l^{\prime }l}^{LR}$ is the
long-range potential that is operative in this channel.  $%
C_{l^{\prime }l}p^{\prime }{}^{l^{\prime }}p^{l}f$ represents the contact interaction, where $f$ can be energy or momentum-dependent. First we consider the case $f=1$, which is the simplest contact term in a given partial wave. To relate the $t$-matrix to a physical quantity, a generalized
scattering length for arbitrary angular momenta $l$ and $l^{ \prime}$ can be
defined as~\cite{vald}~$\frac{\alpha _{l^{\prime }l}}{M}=\lim_{k\rightarrow 0}\frac{%
t_{l^{\prime }l}(k,k;E)}{k^{l^{\prime }+l}},$
where for $l^{\prime }=l=0$ the usual definition, $\frac{%
\alpha _{00}}{M}=t_{00}(0,0;0)$, is obtained. Dividing the partial-wave LSE,
Eq.~(\ref{eq:7}), by $p^{\prime }{}^{l^{\prime }}p^{l}$ we obtain 
\begin{equation}
\frac{t_{l^{\prime }l}^{SJ}(p^{\prime },p;E)}{p^{\prime }{}^{l^{\prime
}}p^{l}}=\frac{v_{l^{\prime }l}^{SJ}(p^{\prime },p)}{p^{\prime
}{}^{l^{\prime }}p^{l}}+\sum_{l^{\prime \prime }}\frac{2}{\pi }\frac{M}{%
p^{\prime }{}^{l^{\prime }}p^{l}}\int_{0}^{\Lambda }\frac{dp^{\prime \prime
}\;p^{\prime \prime }{}^{2}\;v_{l^{\prime }l^{\prime \prime
}}^{SJ}(p^{\prime },p^{\prime \prime })\;t_{l^{\prime \prime
}l}^{SJ}(p^{\prime \prime },p;E)}{p_{0}^{2}+i\varepsilon -p^{\prime \prime
}{}^{2}}.  \label{eq:8.5}
\end{equation}

\noindent Since $v_{l^{\prime }l}^{LR}(p^{\prime },p)\sim p^{\prime
}{}^{l^{\prime }}p^{l}$, Eq.~(\ref{eq:8.5}) is general and can
be applied to any partial wave. 

In the following we concentrate on P-waves ($%
l=l^{\prime }=1$). (The corresponding argument for S-waves, in the case that we have the standard LO contact interaction of $\chi$PT with $l=l'=0$, $f=1$, is analogous, but more straightforward, as division by a factor of $p' k$ is not necessary there.) Consider the half-shell and on-shell $t$-matrices
at $E=0$:
\begin{eqnarray}
\lim_{k\rightarrow 0} \left[ \frac{t_{l^\prime l}(p^{\prime },k;0)}{%
p^{\prime }k} \right] &=&\lim_{k\rightarrow 0}\left[ \frac{v^{LR}_{l^\prime
l }(p^{\prime },k)}{p^{\prime }k} +C_{l^\prime l} \right] \cr &+&
\sum_{l^{\prime \prime }}\frac{2}{\pi }M\lim_{k\rightarrow 0} \left[ \frac{1%
}{p^{\prime }k}\int_{0}^{\Lambda }\frac{dp^{\prime \prime }\;p^{\prime
\prime }{}^{2}\;(v^{LR}_{l^\prime l^{\prime \prime} }(p^{\prime },p^{\prime
\prime })+ C_{l^\prime l^{\prime \prime}}p^{\prime }p^{\prime \prime
})\;t_{l^{\prime \prime }l}(p^{\prime \prime },0;0)}{-p^{\prime \prime
}{}^{2}} \right]  \label{eq:9} \\
\lim_{k\rightarrow 0}\left[ \frac{t_{l^{\prime }l}(k,k;0)}{kk} \right]
&=&\lim_{k\rightarrow 0}\left[ \frac{v^{LR}_{l^\prime l }(k,k)}{kk}%
+C_{l^\prime l}\right] \cr &+& \sum_{l^{\prime \prime }}\frac{2}{\pi }%
M\lim_{k\rightarrow 0}\left[ \frac{1}{kk}\int_{0}^{\Lambda }\frac{dp^{\prime
\prime }\;p^{\prime \prime }{}^{2}\;(v^{LR}_{l^\prime l^{\prime \prime}
}(k,p^{\prime \prime })+C_{l^\prime l^{\prime \prime}}kp^{\prime \prime })\;
t_{l^{\prime \prime }l}(p^{\prime \prime },0;0)}{-p^{\prime \prime }{}^{2}} %
\right].  \label{eq:10}
\end{eqnarray}
\noindent Subtracting Eq.~(\ref{eq:10}) from Eq.~(\ref{eq:9}) and
multiplying both sides by $p^{\prime }$ cancels the unknown $C_{l^\prime
l}$:
\begin{eqnarray}  
\lim_{k\rightarrow 0} \left[ \frac{t_{l^{\prime }l}(p^{\prime },k;0)}{k} %
\right]  & =& \frac{\alpha _{11}}{M}p^{\prime }+\lim_{k\rightarrow 0} \left[%
\frac{v^{LR}_{l^\prime l}(p^{\prime },k)}{k} \right] -p^{\prime}\lim_{k%
\rightarrow 0} \left[ \frac{v_{l^\prime l}^{LR}(k,k)}{ k^{2}} \right] 
\nonumber \\
&-&\sum_{l^{\prime \prime }}\frac{2}{\pi }M\int_{0}^{\Lambda }
dp^{\prime \prime }\; \left[v^{LR}_{l^\prime
l^{\prime \prime} }(p^{\prime },p^{\prime \prime })-\lim_{k\rightarrow 0}[%
\frac{v^{LR}_{l^\prime l^{\prime \prime} }(k,p^{\prime \prime })}{k} ]
p^{\prime } \right]\lim_{k\rightarrow 0}[\frac{\;t_{l^{\prime \prime
}l}(p^{\prime \prime },k;0)}{k}]. \label{eq:11} 
\end{eqnarray}
Here we have used that for P-waves $%
\lim_{k\rightarrow 0}\left[ \frac{t_{11}(k,k;0)}{kk}\right] =\frac{\alpha
_{11}}{M}$. The above limits are well-defined. The only unknown
in Eq.~(\ref{eq:11}) is $\lim_{k\rightarrow 0}\left[ \frac{t_{l^{\prime
}l}(p^{\prime },k;0)}{k}\right],$ which can be solved by
standard techniques.

The next step is to apply
the same idea again to obtain $\frac{t_{l^{\prime }l}(p,p^{\prime
};0)}{p}$ and hence, $t_{l^{\prime }l}(p,p^{\prime };0)$. We then proceed to
calculate the on-shell $t$-matrix and the phase shifts using resolvent
identities that connect the operator $t(E)$ to the operator $t(0)$. Those
details are laid out in Refs.~\cite{AP04,Ya08,Fr99,Ti05}.

Next, we consider an energy-dependent contact term in S-waves. We take  
$fC_{00}=\lambda +\gamma E$. This is the contact term up to NLO and NNLO in $\chi$ET for the $^1$S$_0$ channel. To simplify the presentation, we adopt the following operator notation for
the LSE 
\begin{equation}
t(E)=\lambda +\gamma E+v_{LR}+\left[ \lambda +\gamma E+v_{LR}\right] \
g_{0}(E)\ t(E),  \label{eq:4.4}
\end{equation}%
where $g_{0}(E)$ is the free resolvent of the LSE. Setting $E=0$ in Eq.~(\ref{eq:4.4}) leads to 
$t(0)=\lambda +v_{LR}+\left[ \lambda +v_{LR}\right] \ g_{0}(0)\ t(0),$
which contains only one unknown, $\lambda $. Therefore,
the matrix element $t(p^{\prime },p;0)$ can be obtained from one
experimental datum, here the NN scattering length $a_{0}.$
After applying the same idea to obtain $t(E^{\ast})$ from the phase shifts at an arbitrary energy, and using properties of the LSE to eliminate $\gamma$, we have
\begin{eqnarray}
t(E)+t(0)[g_{0}(0)-g_{0}(E)]t(E)+\frac{E}{E^{\ast }}\Big\{t(0)-\left[
1+t(0)g_{0}(0)\right] \alpha t(E^{\ast })\Big\}\ g_{0}(E)t(E)\cr=\left( 1-%
\frac{E}{E^{\ast }}\right) t(0)+\frac{E}{E^{\ast }}\Big[1+t(0)g_{0}(0)\Big]%
\alpha t(E^{\ast }),\label{eq:4.96}
\end{eqnarray}
where $\alpha \equiv \lbrack 1+t(E^{\ast })g_{0}(E^{\ast })]^{-1}$. With $t(0)$
and $t(E^{\ast })$ known, Eq.~(\ref{eq:4.96}) is an equation for $t(E)$ and can be solved by standard methods with detail
given in Ref.~\cite{Ya09s}.

Finally, for (more complicated) contact terms such as

(A) $\lambda +C_{2}(p^{2}+p^{\prime 2})$;     (B) $\left( 
\begin{array}{cc}
\lambda +C_{2}(p^{2}+p^{\prime 2}) & \lambda _{t}\ p^{\prime 2} \\ 
\lambda _{t}\ p^{2} & 0%
\end{array}%
\right)$;    (C) $\left( 
\begin{array}{cc}
\lambda +\gamma E & \lambda _{t}\ p^{\prime 2} \\ 
\lambda _{t}\ p^{2} & 0%
\end{array}%
\right), $

\noindent we can solve the problem by combining the above methods, i.e., use the first subtraction to eliminate $\lambda$, and then relate $t(E^{\ast })$ to $t(E)$.
For coupled channels, we can apply the idea of dividing $p^{\prime
}{}^{l^{\prime }}p^{l}$ in LSE to eliminate $\lambda _{t}\ p^{2}$.
However, in the case of the
momentum-dependent S-wave contact terms, we need to perform one fitting to eliminate the
unknown constant $C_{2}$. The inputs needed are:

for case (A); $a_{0}$, and an additional data to perform the fitting;

for case (B); $a_{0}$, $\alpha_{20}$, and an additional data to perform the
fitting; and

for case (C); $a_{0}$, $\alpha _{20}$ and phase shift at an arbitrary energy 
$\delta (E^{\ast })$.~\footnote{The only restriction on $E^{\ast }$ is that it
must be within the domain of validity of our theory.
}

\section{Results and discussion}
\label{sec-result}

In this section we present our results in P-waves and S-waves to demonstrate the following:

(1). Our subtractive
renormalization scheme generates results equivalent to the conventional
\textquotedblleft fitting\textquotedblright\ method, with a direct input of
physical observables. 

(2). The energy-dependent contact term produces phase
shifts that oscillate with respect to $\Lambda $. 

(3). Whether a contact
term (or contact terms) is needed for a cutoff independent result is
exactly determined by the (coordinate-space) singularity structure of the
potential as $r\rightarrow 0$. 

(4). Cutoff independence in the phase shift
does not neccessarily mean the results are
renormalization-point independent. Both properties are necessary conditions for
successful renormalization. 

(5). In general, there is a highest cutoff $%
\Lambda _{c}\approx 1$ $(2)$ GeV in the LSE one can adopt for the NNLO DR\ (SFR) TPE, before the results start to exhibit problems.
\begin{figure}
\begin{center}
\includegraphics[width=14cm,height=6.8cm]{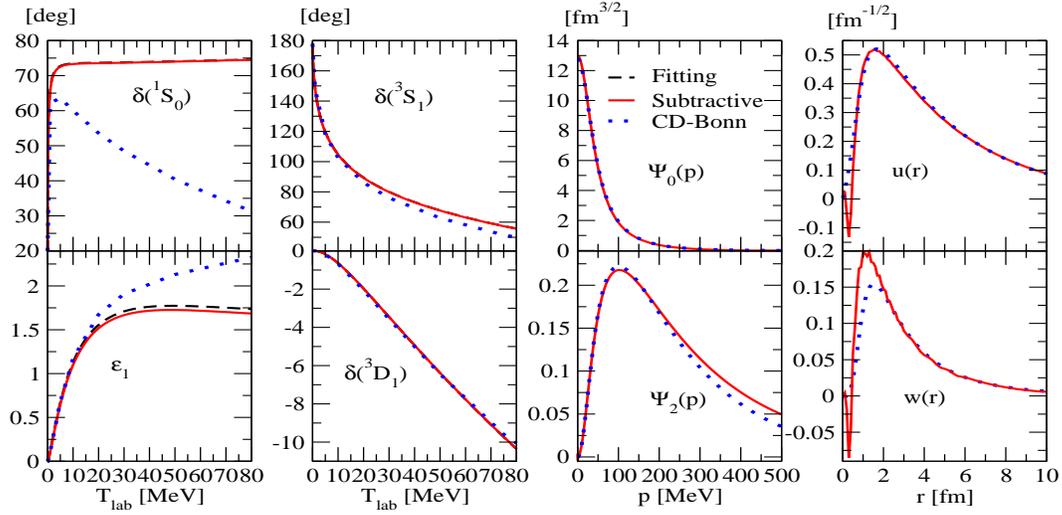}
\end{center}
\caption{(Color online) The comparison of two renormalization methods for the lowest
NN $^1$S$_0$ $\&$ $^3$S$_1$-$^3$D$_1$ phase shifts (left two panels) and the bound-state wavefunctions (right two panels). $\protect \psi_0$(p) (u(r)) is the $^3$S$_1$ wavefunction and
 $\protect\psi_2$(p) (w(r)) denotes the $^3$D$_1$ wave in momentum (coordinate) space. Here $\Lambda=50$ GeV is used. 
The dotted lines indicate the corresponding results obtained with
the CD-Bonn potential~\cite{CDBONN}.}
\label{pos1}
\end{figure}

To show point (1), we compare the LO S-wave phase shift obtained from the
conventional \textquotedblleft fitting\textquotedblright\ method to our
subtractive scheme in the left-hand side of Fig. \ref{pos1}. Here the potential is the
one-pion-exchange (OPE) plus a constant contact term. Fig. \ref{pos1} shows that the results obtained by these two methods agree with each other within a
relative difference of 2\%.  (Due to numerical effects this difference is amplified in $\epsilon _{1}$. $\epsilon _{1}$ is well known for its sensitivity of the
value of the unknown constant in the contact term, and we only adjusted $%
a_{0}$ up to certain precision when
performing the \textquotedblleft fitting\textquotedblright\
method.)
We have verified that the off-shell t-matrices we obtain also agree with those
obtained from the conventional \textquotedblleft fitting\textquotedblright\
method to the same accuracy. Therefore, our subtraction
method is as valid as the conventional fitting method. Our
method can be applied to bound-state calculations too. The right-hand side of Fig. \ref%
{pos1} shows deuteron wavefunctions obtained from our subtractive
method, which are quite close to those obtained from the CD-Bonn potential.
\vspace*{3mm}
\begin{figure}[ht]
\begin{minipage}[h]{8.5cm}
 \includegraphics[width=76mm]{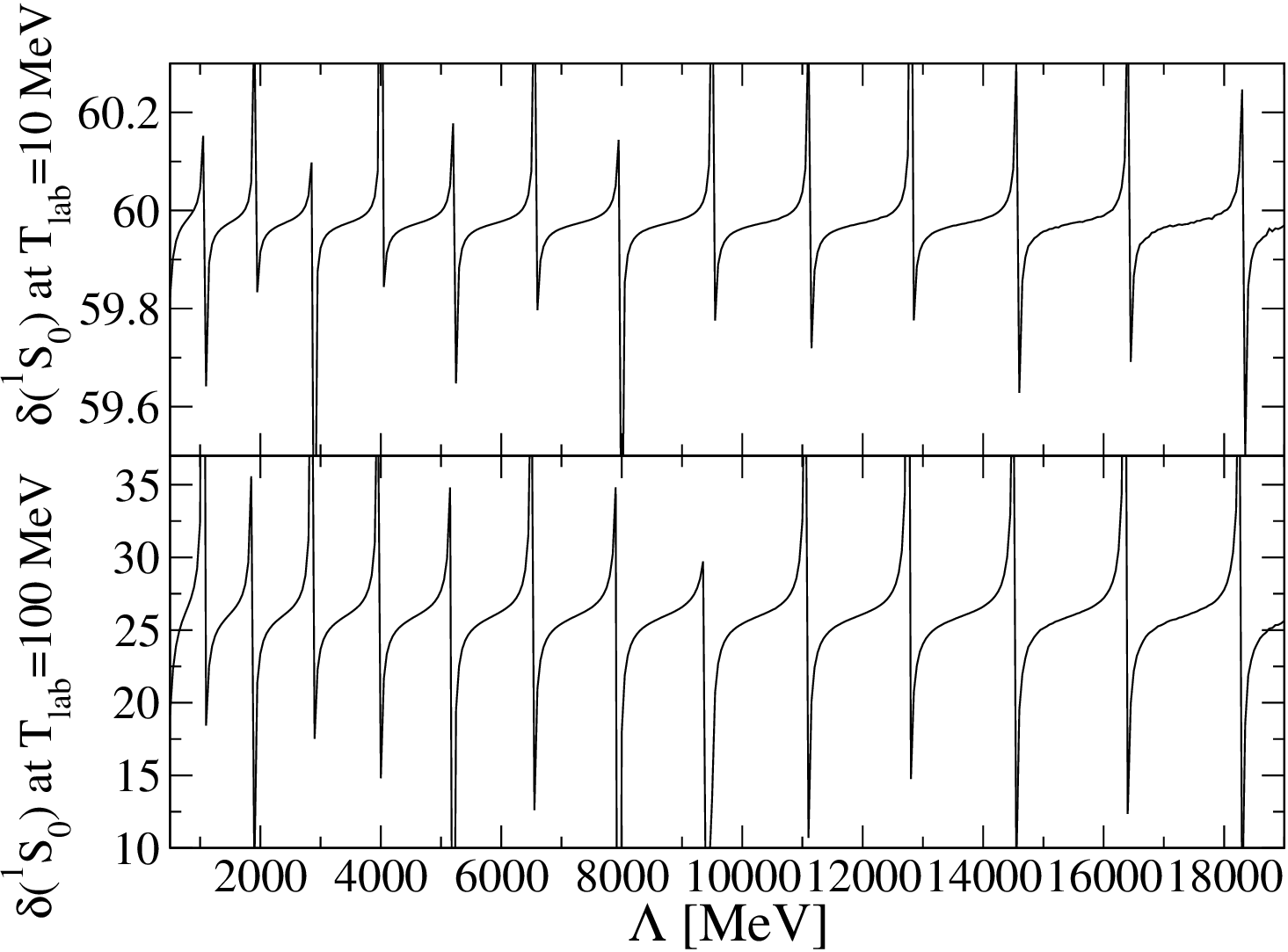}
\label{fig2}
\end{minipage}~%
\begin{minipage}[h]{6.5cm}
{\small {\bf Figure 2: } The $^1$S$_0$ NN phase shift at $T_{lab}=10$ (upper panel) and 100 (lower panel) MeV as a function of $\Lambda$. The results are obtained using the DR
NNLO TPE with an energy-dependent contact term via our subtractive renormalization.
}
\end{minipage}
\end{figure}

For point (2), we associate the DR NNLO TPE with the
energy-dependent contact term and plot the $^{1}$S$_{0}$ phase
shifts versus $\Lambda$ in Fig. 2. The phase shifts show an oscillatory behvaior as a function of $\Lambda$. A similar oscillatory feature is observed in the 
${}^3$S$_1-{}^3$D$_1$ channel. (We use $a_{0},$ 
$\delta (E^{\ast })$ and $\alpha _{20}$ (for the triplet) as the
input to generate the results with $E^{\ast }=1.4$ $(10)$ MeV for the singlet (triplet) channel.)
 This phenomenon is caused by
the resonance state created by the energy-dependent potential. 
We emphasise that
the first place where the phase shifts diverge is at cutoff $\Lambda \approx
1$ $(1.2)$ GeV for the singlet (triplet) channel.
\begin{table}
\begin{minipage}[h]{9cm}
\begin{tabular}{|l|cccc|} \hline
    & $^1$P$_1$ & $^3$P$_0$ & $^3$P$_1$ & $^3$P$_2$ \\
\hline\hline
OPE    & U &  R & U  & R \\
NLO (DR)     & U &  U & R  & R \\
NNLO (DR)  & R &  R & R  & * \\
NLO (DR) + NNLO (SFR)       & U &  U & R  & R \\
NNLO (SFR)       & U &  U & R  & R \\
\hline\hline
\end{tabular}
\label{t1}
\end{minipage}~%
\begin{minipage}[h]{5.5cm}
{\small {\bf Table~I: }
Singularity structure of the long-range potentials $v_{11}^{LR}$. Here ``U" (``R") means that $v_{11}^{LR}$ is repulsive (attractive) at $r\rightarrow 0$. The * indicates that both eigenpotentials in the ${}^3$P$_2-{}^3$F$_2$ channel are attractive.}
\end{minipage}
\end{table}

\begin{figure}
\begin{center}
 \includegraphics[width=14cm,height=7cm]{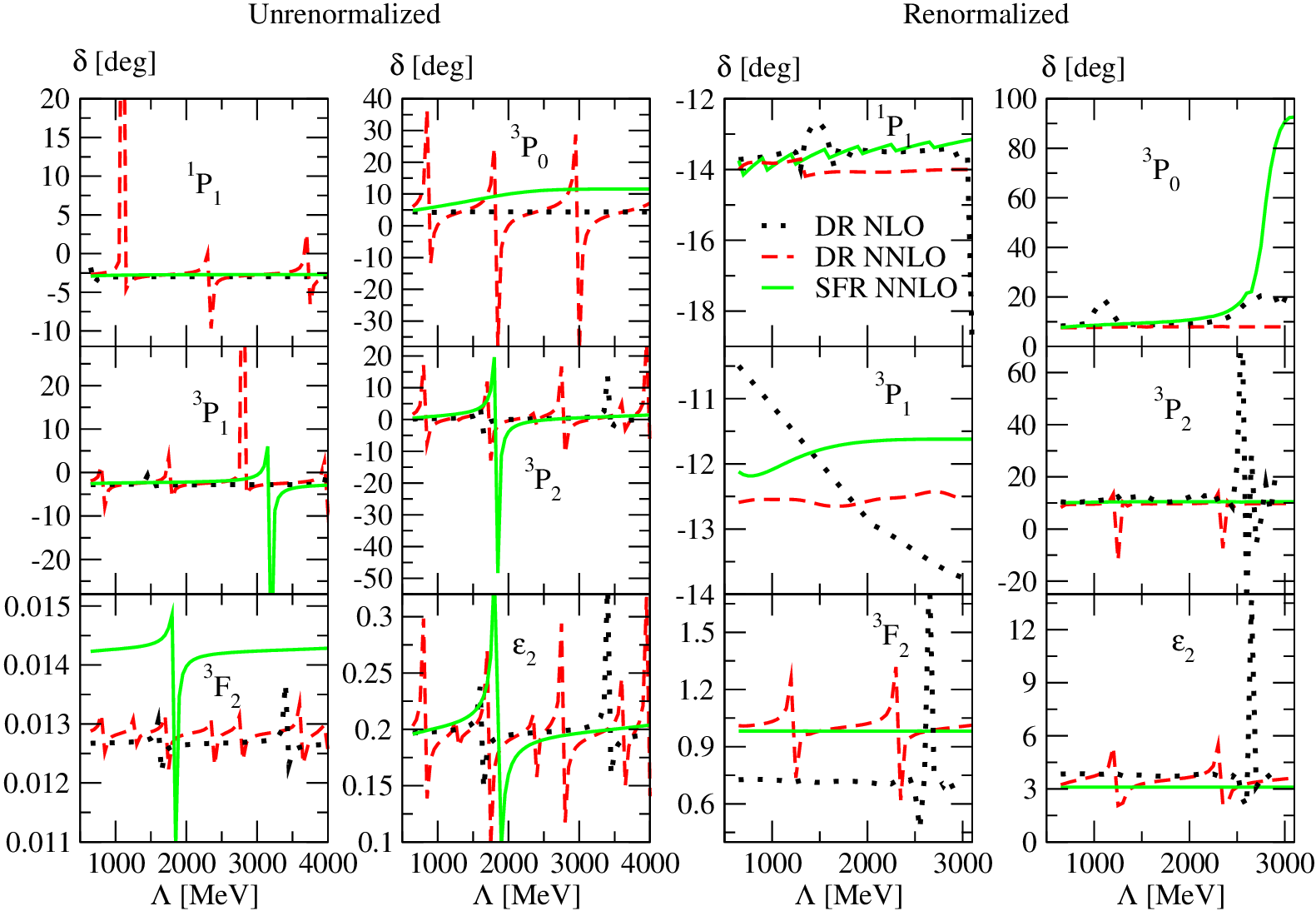}
\end{center}
{\small {\bf Figure 3:} (Color online) The un-renormalized v.s. renormalized NN P-wave phase shifts at $T_{lab}=10$~MeV (un-renormalized) and $100$ MeV (renormalized) as a function of $\Lambda$ for various $\chi$PT potentials: DR NLO, black dotted line; DR NNLO, red dashed line; SFR NNLO, solid green line. For the renormalized case, the input $\alpha _{11}^{SJ}$ were adjusted at each cutoff to give the best
fit to the Nijmegen analysis~\cite{nnonline} in the region $T_{lab}< 100$~MeV.
}
\end{figure}

(3) involves the short distance ($r\rightarrow 0$) behavior of
$v_{l^{\prime }l}^{LR}$ in the coordinate space. We calculate the $r\rightarrow 0$ behavior
analytically for various P-waves potentials and list them in Table~I. At the same time, we plot the un-renormalized v.s. renormalized
phase shifts at $T_{lab}=10$ $(100)$~MeV for TPE up to DR NLO, DR NNLO
and the SFR TPE up to NNLO in Fig. 3. (Here, and throughout this paper, we adopt an intrinsic cutoff $\widetilde{\Lambda}=800$ MeV for the SFR TPE.) For the renormalized cases, the contact term has the form $C_{l^{\prime
}l}^{SJ}p^{\prime }{}^{l^{\prime }}p^{l}$. Comparing the un-renormalized v.s. renormalized case indicates
whether a contact term is needed for the phase shift to be stable with
respect to $\Lambda $. This is exactly determined by the $r\rightarrow 0$ structure listed in Table I. If the
potential is singular and attractive for $r\rightarrow 0$ (denoted
as\textquotedblleft R\textquotedblright\ in Table I), then the
contact term is required. If it is not (\textquotedblleft
U\textquotedblright\ in Table I) then the phase shifts will have a
stable $\Lambda \rightarrow \infty $ limit even in the absence of a contact
term (see also
Refs.~\cite{PVRA06A,PVRA06B,Be01}). The ${}^{3}$P$_{2}-{}^{3}$F$_{2}$
channel for DR NNLO is a special case, since the coupled-channels potential
has two attractive singular eigenpotentials in the $r\rightarrow 0$ limit,
and so one subtraction is not sufficient to make phase shifts independent of $%
\Lambda $ in this channel. 
\begin{figure}[ht]
\begin{minipage}[h]{9.2cm}
 \includegraphics[width=90mm]{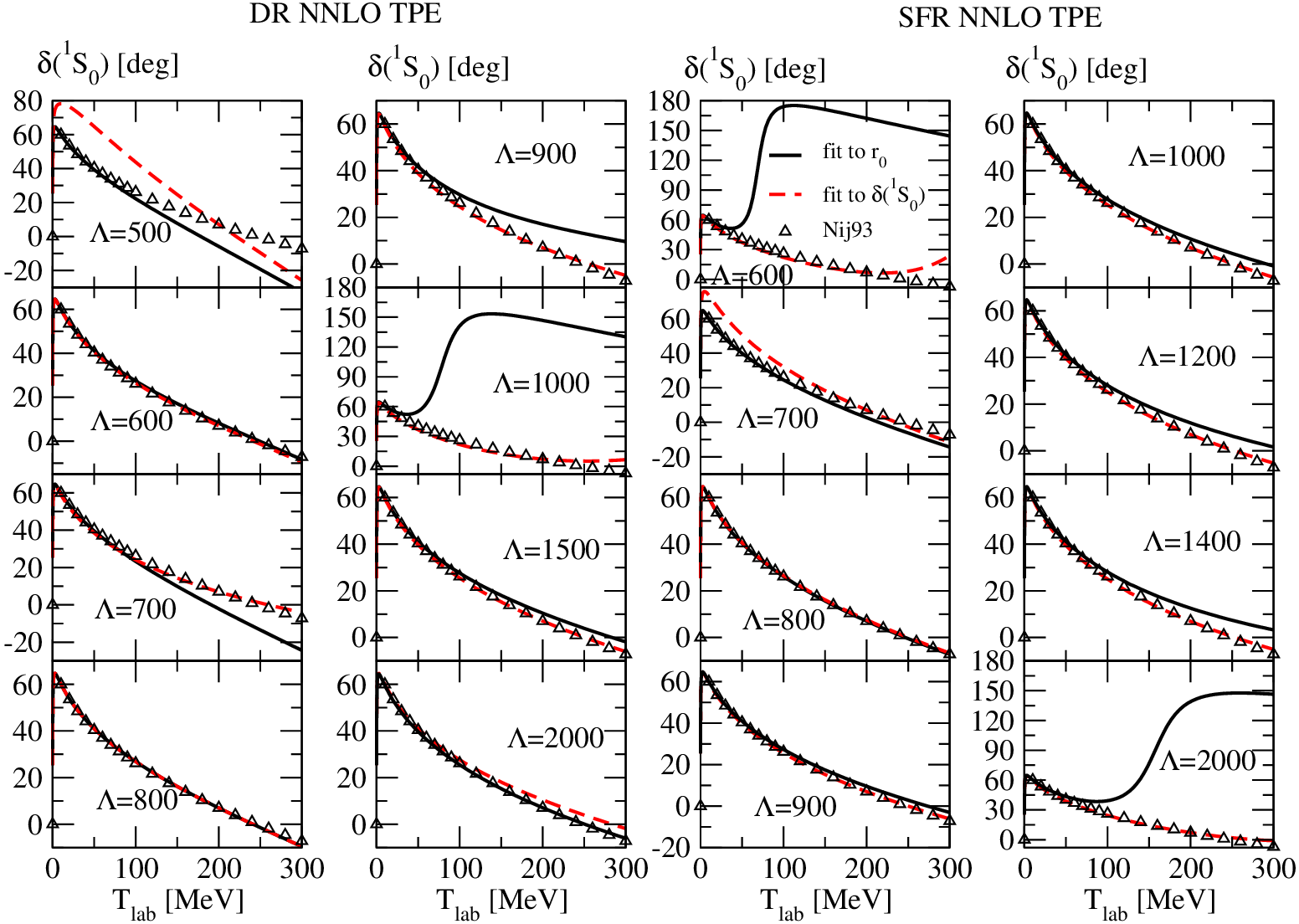}
\label{fig6}
\end{minipage}~%
\begin{minipage}[h]{5.8cm}
{\small {\bf Figure 4}:
(Color online) The $^1$S$_0$ NN phase shift as a function of the lab. kinetic energy for various $%
\Lambda$. The results are obtained with the DR (left two panels) or SFR TPE (right two panels) up to NNLO with momentum-dependent
contact terms. We use $%
a_0=-23.7$ fm as input and then perform a fit to either the effective
range $r_0=2.7$ fm (solid black line) or the phase shift at $%
T_{lab}=200$~MeV (dashed red line). The phase shifts~\cite{nnonline} are denoted by open triangles. }
\end{minipage}
\end{figure}

(4) To see why a cutoff-independent result in the phase shift is not
neccessarily renormalization-point-independent, we plot the 
$^{1}S_{0}$ phase shift obtained with the DR and SFR NNLO TPE along with the
momentum-dependent contact term (denoted as case (A) at the end of the previous
section) in Fig. 4. As mentioned before, in this case
we perform the renormalization by the one-subtraction-plus-one-fitting
procedure. The results obtained by fitting to the effective range $r_{o}$ or to the phase shift at $%
T_{lab}=200$ MeV are shown. One can see that the two different fit
procedures generate different results for the same $\Lambda $. This is
especially visible at $\Lambda $~=~500 and 1000~MeV for the DR NNLO TPE,
where a resonance-like behavior is present in the latter case when $C_{2}$
is fitted to $r_{0}$. For values of $\Lambda $ not close to these
problematic cutoffs the phase shift is almost independent of the
renormalization point. In contrast, for the SFR TPE, the two different fitting procedures lead
to almost the same phase shift for $\Lambda $ between $700-1800$~MeV. By
switching to the SFR TPE, we achieve renormalization-point-independence for
a wider range of $\Lambda$.

\vskip 0.5cm

\begin{figure}[ht]
\begin{minipage}[h]{8.5cm}
 \includegraphics[width=80mm]{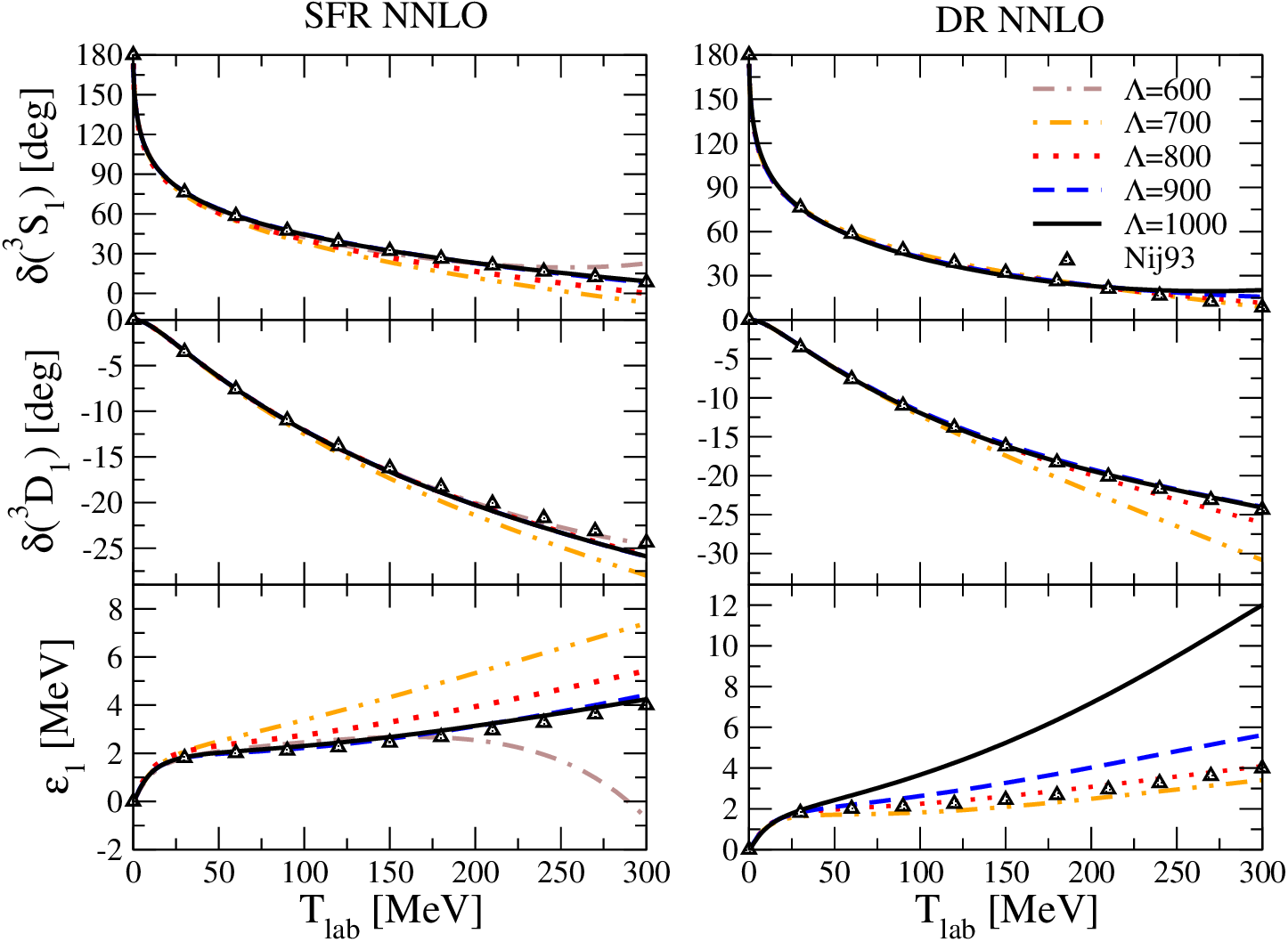}
\label{fig7}
\end{minipage}~%
\begin{minipage}[h]{6.2cm}
{\small {\bf Figure 5: } (Color online) The best fit for the NN $^3$S$_1-^3$D$_1$ phase
shifts as a function of the laboratory kinetic energy for different cutoffs $%
\Lambda$ ranging from 0.6 to 1~GeV. The potentials employed are the SFR NNLO
with a momentum-dependent central part of the contact term. The values of
the Nijmegen phase-shifts~\protect\cite{nnonline} are indicated by the open
triangles.
}
\end{minipage}
\end{figure}

Finally, for point (5), we plot the ${}^{3}$S$_{1}-{}^{3}$D$_{1}$ phase
shifts in Fig. 5. These results are obtained by the SFR or DR TPE up to NNLO plus the momentum-dependent contact term (labeled as (C) in the previous section). For the DR TPE case, the best overall fit already diverges away from the Nijmegen analysis in the mixing angle at $\Lambda=1000$ MeV. This implies that there is a critical cutoff $\Lambda_{c}\sim1$ GeV. Above that we cannot iterate DR NNLO TPE in LSE and obtaine a good fit in the ${}^{3}$S$_{1}-{}^{3}$D$_{1}$ channel. Moreover, as shown in Fig. 4, at this $\Lambda_{c}$ the renormalization-point-independence also breaks down for the DR NNLO TPE in the $^1$S$_0$ channel. Therefore, we conclude that for the DR TPE in S-waves, the highest cutoff one can adopt in the LSE is $\Lambda_{c}\sim1$ GeV. For the P-waves, a detail analysis of the renormalization-point-dependence suggests that $\Lambda_{c}\sim1-1.2$ GeV for the DR NNLO TPE\cite{Ya09p}. As with the S-waves, if the SFR TPE is adopted, then $\Lambda_{c}$ can be extented to 2 GeV before similar problems appear. 


\section{Summary and Conclusions}

\label{sec-con}

We developed a subtractive renormalization scheme for $\chi$ET NN
potentials which allows us to
go to an arbitrarily high cutoff in the LSE.
Our calculations show that the energy-dependent contact term creates scattering resonances and
shallow bound states in S-wave channels once cutoffs larger than 1 GeV are
considered. Momentum-dependent contact terms in the NNLO
potential also has problems at these cutoffs.
We also investigate the singularity structure of the potential and find that the LO conclusion presented in Ref.\cite{NTvK05} holds up to NNLO. Our analysis in S-waves and P-waves shows that the two-pion-exchange potential should
not be inserted in the Lippmann-Schwinger equation and treated
non-perturbatively if cutoffs larger than 1 GeV are employed.



\section*{Acknowledgments}

This work was performed in part under the auspices of the U.~S. Department
of Energy, Office of Nuclear Physics, under contract No. DE-FG02-93ER40756
with Ohio University. We thank the Ohio Supercomputer Center (OSC) for the
use of their facilities under grant PHS206. 



\end{document}